\documentclass[final]{iopart}
\usepackage{iopams,graphicx,texdraw,epsf,bm}

\newcommand{\tb}{\mathring{\tau}}
\newcommand{\ub}{\mathring{u}}
\newcommand{\cb}{\mathring{c}}
\newcommand{\eqref}[1]{(\ref{#1})}
\eqnobysec
\begin{document}
\paper[$O(n)$ $\phi^4$ model with free surfaces in the large-$n$ limit\ldots]
{The  $O(n)$ $\phi^4$ model with free surfaces in the large-$n$  limit: Some exact results for boundary critical behaviour, fluctuation-induced forces and distant-wall corrections}
\author{H.~W.  Diehl and S.~B.\ Rutkevich}
\address{%
  Fakult\"at f\"ur Physik, Universit{\"a}t Duisburg-Essen, D-47048 Duisburg,
  Germany}
\begin{abstract}
The  $O(n)$ $\phi^4$ model on a  slab $\mathbb{R}^{d-1}\times[0,L]$ bounded by free surfaces is studied for $2<d<4$ in the limit $n\to\infty$. The self-consistent potential $V(z)$ which the exact  $n\to\infty$ solution of the model involves is analysed by means of boundary operator expansions. Building on the known exact $n\to\infty$ solution for $V(z)$ in the semi-infinite case $L=\infty$ at the bulk critical point, we exactly determine two types of corrections to this potential: (i) those linear in the temperature scaling field  $t$ at  $L=\infty$, and (ii) the leading $L$-dependent (distant-wall) corrections at the critical point. From (i) exact analytical results at $d=3$ are obtained  for the leading temperature singularity of the excess surface free energy and the implied asymptotic behaviours of the scaling functions $\Theta_3(x)$ and $\vartheta_3(x)$ of the residual free energy $f_{\rm res} =L^{1-d}\,\Theta_d(tL)$ and the critical Casimir force $\beta\mathcal{F}_{\rm C}(T,L)=L^{-d}\,\vartheta_d(tL)$ in the limit $x\to 0\pm$. The second derivative $\vartheta_3''(0)$ is computed exactly.
\end{abstract}

\section{Introduction}
Exactly solvable models have played an important role in the development of the theory of phase transitions and critical phenomena (see e.g.\ \cite{Bax07,Mus10}). The mathematically reliable results they produced have provided helpful benchmarks for approximation schemes. Furthermore,  such exact solutions have repeatedly led to surprises, predicting unexpected behaviours. Familiar examples are  Onsager's celebrated solution of the two-dimensional Ising model \cite{Ons44}, which rigorously proved the breakdown of the Landau theory of second-order phase transitions (see e.g.\ \cite[Kap.~XIV]{LL70}), and Baxter's exact solution of the square lattice  eight-vertex  model in zero field \cite{Bax71}, which established the possibility of continuously varying critical indices. 

One familiar class of models that lend themselves to exact solutions are $n$-vector models in the limit $n\to\infty$ \cite{Ma76,MZ03}. In this paper we shall be concerned with $d$-dimensional $n$-component $\phi^4$ models on the $d$-dimensional half-space $z\ge0$ bounded by a single planar ($d-1$)-dimensional hyper-surface at $z=0$ and a film bounded by a pair of such hyper-surfaces at $z=0$ and  $z=L$, where we assume the Hamiltonians of these models  to possess $O(n)$ symmetry in the absence of magnetic fields. 

The analogous case of such models obeying periodic, rather than free, boundary conditions is much easier to handle. In fact, the spherical model that is equivalent to the limit $n\to\infty$ of the $n$-component $\phi^4$ model with periodic boundary conditions has been solved exactly for $d=3$, even in the presence of a magnetic field \cite{Dan96,Dan98}. This gave the scaling functions of the $L$-dependent part of the free energy  per area of the bounding surfaces and of the implied fluctuation-induced (critical Casimir) force\footnote{For background on critical Casimir forces, see references \cite{Kre94,BDT00,Gam09}.} \cite{Kre94,BDT00,Gam09} in closed analytical form. The simplicity of periodic boundary conditions is due to the fact that translation invariance is preserved even along the direction normal to the bounding surfaces ($z$-direction). The presence of free surfaces causes a breakdown of translation invariance along this direction. This implies that the $n\to\infty$ limit of such models leads to modified spherical models which involve separate constraints for each layer $z$ and hence $z$-dependent Lagrange multipliers \cite{Kno73}. To determine the $n\to\infty$ limit  of the free energy and other quantities, the eigenvalues and eigenfunctions of a one-dimensional Schr\"odinger equation must be solved together with a self-consistency condition for the associated potential $V(z)$ \cite{DGHHRS12,BDS10}. Unfortunately, an exact solution of this self-consistency problem at $2<d<4$ is not known in closed analytical form except for the semi-infinite ($L=\infty$) case precisely at the bulk critical temperature $T_{\rm{c}}$ \cite{BM77a,BM77c}. 

Building on this exact solution due to Bray and Moore, we will aim at its extensions  for finite temperature deviations $\tau\propto (T-T_{\rm{c}})/T_{\rm{c}}$ from the critical point and for finite film thickness $L$. The organisation of the paper is as follows. In the next section, we define the model, give the self-consistency problem to which its exact solution in the limit $n\to\infty$ leads, recollect its bulk solution, analyse the scaling form of the self-consistent potential by means of the boundary operator expansion (BOE), and recall Bray and Moore's exact solution for the semi-infinite case at the bulk critical point. In section~\ref{sec:tcorr} we compute the leading temperature-dependent correction to the self-consistent potential for the semi-infinite case. The implications for the excess free energy and the scaling functions of the residual free energy and the critical Casimir force  are worked out in section~\ref{sec:conseq}. Section~\ref{sec:conc} contains a brief summary of the obtained exact results for the $(d=3)$-dimensional case and concluding remarks. Finally, there is an appendix in which the  calculation of the bulk free energy density is described for our cutoff scheme.

\section{The model and its large-$n$ solution}\label{sec:modsol}

\subsection{The model}

We consider a classical model for an $n$-component order-parameter field $\bm{\phi}(\bm{x})=(\phi_\alpha(\bm{x}))$ described by the Hamiltonian
\begin{eqnarray}\label{eq:Ham}
\mathcal{H}[\bm{\phi}]&=&\int_0^L\rmd{z}\int\rmd^{d-1}y\left[\frac{1}{2}(\nabla\bm{\phi})^2+\frac{\tb}{2}\phi^2+\frac{\ub}{4!n}\phi^4\right]\nonumber\\&&\strut 
+\frac{\cb}{2}\int\rmd^{d-1}y\left[ \phi^2\big|_{z=0}+\phi^2\big|_{z=L}\right],
\end{eqnarray}
where $\alpha=1,\ldots,n$ labels the components of $\bm{\phi}$, $\phi$ means its absolute value $|\bm{\phi}|$ and $(\nabla\bm{\phi})^2$ is  short-hand for $\sum_{\alpha=1}^n(\nabla\phi_\alpha)^2$. We have decomposed the position vector $\bm{x}=(\bm{y},z)$ into components $\bm{y}\in\mathbb{R}^{d-1}$ and $z\in [0,L]$ parallel and perpendicular to the $z=0$ plane, respectively. The first term in equation~\eqref{eq:Ham} is an integral over the standard $\phi^4$ bulk density, where we have included a factor of $1/n$ in the $\phi^4$ interaction term to make the limit $n\to\infty$ well-defined. The second term consists of contributions localised on the boundary planes $z=0$ and $z=L$, whose interaction constants we have chosen to take the same value $\cb$. As is well known \cite{Die86a}, these boundary terms give rise to the Robin boundary conditions 
\begin{equation}
\partial_n\bm{\phi}=\cb\,\bm{\phi},\quad z=0,L,
\end{equation}
where $\partial_n$ denotes the derivative along the inner normal $\bm{n}$. Unless otherwise stated, we restrict $d$ to values $2<d<4$ between the lower critical dimension $d_*=2$ and the upper one $d^*=4$.

Let $\bm{p}$ be the wave vector conjugate to the variable $\bm{y}$. Translation invariance along the $\bm{y}$-direction implies that  correlation functions of two fields at positions $\bm{x}=(\bm{y},z)$ and $\bm{x}'=(\bm{y}',z')$ have the form $f(\bm{y}-\bm{y}';z,z')$ and are diagonal in $\bm{p}$-space. We use a hat to denote their Fourier $\bm{p}$-transform, defining
\begin{equation}
\hat{f}(\bm{p};z,z')=\int\rmd^{d-1}y\,f(\bm{y};z,z')\,\rme^{-\rmi\bm{p}\cdot\bm{y}}.
\end{equation}
Following Bray and Moore \cite{BM77a,BM77c}, we shall use an ultraviolet cutoff $\Lambda$ to restrict integrations over parallel momenta $\bm{p}$ to values with $p\equiv|\bm{p}|\le\Lambda$. 

Consider the two-point cumulant function
\begin{equation}\label{eq:sceG2}
G^{(2)}_{\alpha\beta}(\bm{y}-\bm{y}';z,z')=\langle\phi_\alpha(\bm{y},z)\,\phi_\beta(\bm{y}',z')\rangle-\langle\phi_\alpha(\bm{y},z)\rangle\langle\phi_\beta(\bm{y}',z')\rangle.
\end{equation}
From  extensions of the Mermin-Wagner theorem \cite{MW66,MW94} it is  known that when $d\le 3$ the $O(n)$ symmetry cannot be spontaneously broken for finiteness thickness $L$.  Hence this symmetry is preserved if $L<\infty$ or $L=\infty$ but $T\ge T_{\rm{c}}$.  Under these conditions $G_{\alpha\beta}^{(2)}$ can be written as
\begin{equation}\label{eq:G2form}
G^{(2)}_{\alpha\beta}(\bm{y};z,z')=\delta_{\alpha\beta}\,G(\bm{y};z,z').
\end{equation}

\subsection{Large-$n$ self-consistency problem}
In the limit $n\to\infty$ the Fourier transform of the function on the right-hand side of equation~\eqref{eq:G2form} satisfies the differential equation
\begin{equation}\label{eq:sceG2}
[-\partial_z^2 +p^2+V_d(z|L,\tau)]\hat{G}(\bm{p};z,z'|L,\tau)=\delta(z-z'),
\end{equation}
where the self-consistent potential $V_d(z|L,\tau)$ is a solution to
\begin{equation}\label{eq:sceV}
V_d(z|L,\tau)-\tau=\frac{\ub}{6}\int_{p,\Lambda}^{(d-1)}\left[\hat{G}(\bm{p};z,z|L,\tau)-\frac{1}{2p}\right].
\end{equation}
Here the integral  symbol on the right-hand side stands for
\begin{equation}
\int_{\bm{p},\Lambda}^{(d-1)}\ldots =\int_{|\bm{p}|\le \Lambda}\frac{\rmd^{d-1}p}{(2\pi)^{d-1}}\ldots,
\end{equation}
while
\begin{equation}
\tau=\tb-\tb_{\rm c}
\end{equation}
measures the deviation of $\tb$ from its bulk critical value $\tb_{\rm c}$. The subtracted term within the square brackets of equation~\eqref{eq:sceV} is the  critical bulk term $\hat{G}(\bm{p};\infty,\infty|\infty,0)=(2p)^{-1}$. This tells us that $\tb_{\rm c}$ is given by
\begin{equation}
\tb_{\rm c}=\frac{-\ub}{6}\int_{p,\Lambda}^{(d-1)}\frac{1}{2p}
\end{equation}
when $n=\infty$. 

\subsection{Bulk  solution}
We shall need a number of bulk results for $n=\infty$. It will therefore be helpful to show how they can be recovered from the self-consistency equations~\eqref{eq:sceG2} and \eqref{eq:sceV}.  The bulk value of the potential is given by $V_d(\infty|\infty,\tau)$. When $\tau\ge 0$, it  corresponds to the inverse of the bulk susceptibility $\chi_{\rm{b}}$, Writing it as
\begin{equation}\label{eq:Vbulk}
V_d(\infty|\infty,\tau\ge 0)=m^2=\xi^{-2},
\end{equation}
we introduce a ``mass'' $m$ and the second-moment bulk correlation length 
\begin{equation}
\xi=\xi_+(\tau/\Lambda^2)^{-\nu}=m^{-1}.
\end{equation}

To compute the non-universal amplitude for $\tau>0$, $\xi_+$, we need the integral
\begin{eqnarray}\label{eq:Vbres}\fl
\int_{\bm{p},\Lambda}^{(d-1)}\left[\frac{1}{2\sqrt{p^2+m^2}}-\frac{1}{2p}\right]&=&
\Lambda ^{d-2}\,K_{d-1}\Bigg[\frac{{}_2F_1\big(\frac{1}{2},\frac{d-1}{2};\frac{d+1}{2};-\frac{\Lambda ^2}{m^2}\big)}{2(d-1) m/\Lambda}-\frac{1}{2 (d-2)}\Bigg]\nonumber\\
&=&-A_d\,m^{d-2}+w_d\,\Lambda^{d-4}\,m^2 +\Lambda^{d-2}\,\Or(m^4/\Lambda^4),
\end{eqnarray}
where ${}_2F_1(\alpha,\beta;\gamma;z)$ is the hypergeometric function. The factor $K_{d-1}$, resulting from the angular integrations, is given by
\begin{equation}\label{eq:Kd}
K_d\equiv 2 (4\pi)^{-d/2}/\Gamma(d/2).
\end{equation}
The coefficient $A_d$ is known to be independent of the chosen regularization scheme (i.e., it is universal) \cite{MZ03} and takes the value
\begin{equation}
A_d\equiv-(4\pi)^{-d/2}\,\Gamma[1-d/2).
\end{equation}
By contrast, the coefficient $w_d$ is non-universal; it does depend on the regularization scheme and can even be negative for some of them, such as lattice regularizations \cite{MZ03,DGHHRS12}. For our type of regularization, one has
\begin{equation}
w_d\equiv\frac{K_{d-1}}{4(4-d)}\;\mathop{=}^{d=3}\frac{1}{8\pi}.
\end{equation}

Upon substituting the second line of  equation~\eqref{eq:Vbres} into equation~\eqref{eq:sceV}, with $L$ and $z$ set to infinity, and introducing the dimensionless coupling constant
\begin{equation}
u=\ub/\Lambda^{4-d},
\end{equation} we arrive at
\begin{equation}\label{eq:taum}
\frac{\tau}{\Lambda^2}\approx\frac{u}{6}\,A_d\,\Big(\frac{m}{\Lambda}\Big)^{d-2}\left[1+\frac{6}{A_d}\Big(\frac{1}{u}-\frac{w_d}{6}\Big) \Big(\frac{m}{\Lambda}\Big)^{4-d}\right].
\end{equation}
From this result we can read off the well-known $n=\infty$ values 
\begin{equation}
\nu=(d-2)^{-1},\quad \omega=4-d,
\end{equation}
of the correlation-length exponent $\nu$ and the correction-to-scaling exponent $\omega$, as well as the non-universal amplitude
\begin{equation}\label{eq:xiplus}
\xi_+=\Lambda^{-1}\Big(\frac{u
   A_d}{6}\Big)^{\frac{1}{d-2}}\;\mathop{=}^{d=3}\frac{u}{24\pi}\,\Lambda^{-1}.
\end{equation}
Furthermore, we see that the leading corrections to scaling vanish when $u$ takes the special value 
\begin{equation}\label{eq:ustar}
u^*=6/w_d\mathop{=}^{d=3}48\pi,
\end{equation}
which is positive for our cutoff scheme and hence may be identified as a fixed-point value \cite{MZ03}. 

In our subsequent analysis it will be convenient to absorb the amplitude $\xi_+\Lambda$ in the dimensionless linear temperature scaling field $t$ by defining
\begin{equation}\label{eq:tdef}
t\equiv\frac{6}{A_d\,u}\,\frac{\tau}{\Lambda^2}.
\end{equation}
This variable can be used both above and below the bulk critical temperature $T_c$. For either sign, i.e.\ for $t=\pm|t|$, it is related to the bulk correlation length $\xi(|t|)$ in the high-temperature phase via $t=\pm |\Lambda\,\xi(\pm t)|^{-1/\nu}$.

\subsection{Scaling form of the self-consistent potential and boundary-operator expansion}

We are interested to determine asymptotic large-length-scale solutions of the self-consistency equations~\eqref{eq:sceG2} and \eqref{eq:sceV}. To this end it is useful to take advantage, as much as possible,  of the available information on the scaling form of such solutions for $\tau\ne0$ and $L\le\infty$. In order to eliminate corrections to scaling, we set the coupling constant $u$ to its fixed-point value~\eqref{eq:ustar}. Unless otherwise stated, we also assume that  the dimensionless surface-enhancement variable $c\equiv \cb/\Lambda$ is set to its fixed-point value $c^*_{\rm{or}}=\infty$ associated with the ordinary transition \cite{Die86a,Die97}, so that  Dirichlet boundary conditions hold on long length scales and corrections to scaling due to the irrelevant surface scaling field $\propto c^{-1}$ are absent.

The right-hand side of \eqref{eq:sceV} is proportional to the deviation of the local energy density from its bulk value at criticality. The energy density scales as $m^{d-1/\nu}$, where the exponent $d-1/\nu=(1-\alpha)/\nu$ simplifies to $2$ when  $n=\infty$. Hence the potential in equation~\eqref{eq:sceV} can be written as
\begin{equation}
V_d(z|L,\tau)=z^{-(d-1/\nu)}\,\Upsilon_d(z/\xi,z/L)
\end{equation}
on  long length-scales, where $\Upsilon_d$ is a dimensionless function. In a renormalization-group (RG) approach 
it would be a fixed-point property and hence universal, since we assumed irrelevant scaling fields such as $u-u^*$ and $1/c$ to vanish. Specifically, at the bulk critical point and infinite thickness $L$, the scaling function $\Upsilon_d$ takes a universal number $\Upsilon_d(0,0)$. Upon solving the self-consistency equations~\eqref{eq:sceG2} and \eqref{eq:sceV} for the semi-infinite case at bulk criticality, Bray and Moore \cite{BM77a,BM77c} determined this universal number for the cases of the ordinary transition with $2<d<4$ and the special transition with $3<d<4$, obtaining the results\footnote{To investigate the case of the special transition, one must of course allow $\cb$ to take the critical value $\cb_{\rm{sp}}$ that it has at the corresponding multi-critical point \cite{Die86a}.}
\begin{equation}\label{eq:BMres}\fl
\Upsilon_d(0,0)=\cases{A_d^{\rm{or}}=\frac{(d-3)^2-1}{4},&$2<d<4$, ordinary transition,\\
A_d^{\rm{sp}}=\frac{(d-5)^2-1}{4},&$3<d<4$, special transition.
}
\end{equation}

Useful additional information about properties of the function $\Upsilon_d$ can be obtained via the BOE \cite{DD83a,Die86a,Die97,Car90b,EKD93,MO95}. For a scaling operator $\mathcal{O}(\bm{y},z)$ with scaling dimension $\Delta[\mathcal{O}]$ the BOE near the plane $z=0$ reads
\begin{equation}\label{eq:BOE}
\mathcal{O}(\bm{y},z)\mathop{=}_{z\to 0}\sum_j z^{\Delta^{(\rm{s})}_j-\Delta[\mathcal{O}]}\,
C_{\mathcal{O}\/,j}(z/\xi,z/L)\,\mathcal{O}^{(\rm{s})}_j(\bm{y}),
\end{equation}
where $\mathcal{O}^{(\rm{s})}_j$ are surface operators with scaling dimension $\Delta^{(\rm{s})}_j$. 
Since our main interest lies in the $(d{=}3$)-dimensional case, where only the ordinary transition remains, we restrict ourselves to the latter. It is well known that in this case the leading operator contributing to the BOE besides the unity operator $\bm{1}$  is given by the $zz$-component $T_{zz}$ of the stress-energy tensor. Upon applying the BOE to the energy density $\varepsilon=\phi(\bm{y},z)^2/n$ and using the fact that the scaling dimension of $T_{zz}$ is $d$, we conclude that 
\begin{equation}\label{eq:BOEres}
\Upsilon_d(\zeta_1{=}zm,\zeta_2{=}z/L)\mathop{=}_{z\to 0} A_d^{\rm{or}}\left[X_d(\zeta_1,\zeta_2)+\zeta_1^d\,Y_d(\zeta_1,\zeta_2)+\ldots\right],
\end{equation}
where the ellipsis stands for contributions from omitted surface operators. The scaling functions $X_d$ and $Y_d$ are expected to have the limiting behaviours
\begin{eqnarray}\label{eq:BOEXd}
X_d(\zeta_1,0)&\mathop{=}_{\zeta_1\to0}&1+a_1(d)\,\zeta_1^{1/\nu}+a_2(d)\,\zeta_1^{2/\nu}+a_3(d)\,\zeta_1^{3/\nu}\ldots,\label{eq:limUps1} \\
Y_d(\zeta_1,0)&\mathop{=}_{\zeta_1\to0}&b_0(d)+b_1(d)\,\zeta_1^{1/\nu}+b_2(d)\,\zeta_1^{2/\nu}+\ldots,\label{eq:limUps2}\label{eq:BOEYd1}\\
\label{eq:BOEYd2}
Y_d(\zeta_1,\zeta_2)&\mathop{=}_{\zeta_1,\zeta_2\to0}&e_0(d)\,(\zeta_2/\zeta_1)^d+\ldots.\label{eq:limUps3}
\end{eqnarray}
The first term on the right-hand side of equation~\eqref{eq:limUps1} is required by consistency with equation~\eqref{eq:BMres}. The terms involving the coefficients $a_i$ and $b_j$ express the expectation that the  short-distance functions $X_d(\zeta_1,0)$ and $Y_d(\zeta_1,0)$ are regular in $t$ near $t=0$ provided $d\ne 3$. The case $d=3$ deserves special considerations when $n=\infty$ since the regular term $\propto a_3t^3$ is degenerate with the one $\propto b_0t^3$. By analogy with known other cases one might expect that the coefficients $a_3(d)$ and $b_0(d)$ vary near $d=3$ as \cite{CK86}
\begin{eqnarray}\label{eq:a3exp}
a_3(d)&=&\frac{1}{2}\frac{\tilde{b}_{-1}}{d-3}+a_{3,0}+\Or(d-3),\\ \label{eq:b0exp}
\pm b^\pm_0(d)&=& -\frac{1}{2}\frac{\tilde{b}_{-1}}{d-3}+\tilde{b}^\pm_0-a_{3,0}+\Or(d-3),
\end{eqnarray}
where we have added $\pm$ signs to distinguish the cases  $t>0$ and $t<0$.
Unless $\hat{b}_{-1}$ vanishes, $\Upsilon_3(\zeta_1,0)$ would become
\begin{equation}
a_3(d)\zeta_1^{3/\nu}+b^\pm_0(d)\,\zeta_1^d\mathop{\longrightarrow}_{d\to 3}t^3\Big[\tilde{b}^\pm_0+\tilde{b}_{-1}\ln |t|\Big],
\end{equation}
at $d=3$ and hence involve  a logarithmic anomaly. Here, we used the fact that $\nu=1/(d-2)=1+O(d-3)$.

However, this degeneracy of the singular term $\sim |t|^{d\nu}$ and the regular one $\sim t^3$ at $d=3$ occurs also for the bulk free energy density, which does not exhibit such a logarithmic anomaly for $n=\infty$ at $d=3$  and whose  amplitude of the $t^{d\nu}$ term is known to remain finite as $d\to3$ \cite{Bax71,Ma76,MZ03,DGHHRS12,AH75}. Likewise,  the amplitude $a_3(d)$ should remain finite as $d\to 3$, so that the residue $\tilde{b}_{-1}$ vanishes  and no logarithmic anomaly arises at $d=3$.

To see this, note that the term $\sim |t|^{d\nu}$ gives us the leading temperature singularity of the energy density near the surface, i.e., of the local surface energy density \cite{Die86a}. For the case of the ordinary transition, the amplitudes $E_{1,\pm}$ of this singularity $E_{1,\pm}\,| t|^{d\nu}$  for $t\to 0\pm$ have been shown to be proportional to their analogues $A_\pm$ of the leading thermal singularity $A_\pm|t|^{d\nu}$ of the bulk free energy density, where the ratios $E_{1,_+}/E_{1,-}$ and $A_+/A_-$ take the same universal  value \cite{BD94}.  According to \cite{AH75}, one has
\begin{equation}\label{eq:ApAmdne3}\fl
\lim_{n\to\infty}\frac{1}{n}\,\frac{A_+}{A_-}=\frac{ \Gamma(d/2) \Gamma(2-d/2) }{\Gamma\left(\frac{4-d}{d-2}\right)\Gamma \left(\frac{2d-6}{d-2}\right)}\, \left(\frac{2^{3-d}\,\Gamma(3/2)\,\Gamma(d/2)}{\Gamma[(d-1)/2]}\right)^{d/(d-2)}\mbox{ for } 3<d<4
\end{equation}
and
\begin{equation}\label{eq:ApAm}
\lim_{n\to\infty}\frac{A_+}{A_-}=\frac{\pi^2}{4}-1\quad\mbox{for } d=3.
\end{equation}
Both results require calculations to the next order in $1/n$. This is obvious for the first one, equation~\eqref{eq:ApAmdne3}, which is in accordance with the  $n=\infty$ result for the singular part of the bulk free energy  derived in the appendix. In the special case $d=3$ where $2-\alpha$ becomes an integer ($=3$), the identification of the leading singular part is impeded by its interference with regular $t^3$ contributions. As is discussed in \cite{AH75}, ambiguities in the identification of  $A_\pm$ can be avoided by going beyond lowest order in the $1/n$ expansion.

Finally, the contribution  $\propto e_0(d)$ describes the effect of a second far plane on the energy density near the  plane $z=0$ at bulk criticality, i.e., this coefficient is a distant-wall amplitude \cite{Car90b,EKD93,FdG78,RJ82}. Its value can be gleaned from the $n=\infty$ results obtained in \cite{MO95}. Let $B_\varepsilon^{T}(d)/n$ be the BOE expansion coefficient denoted as $B_{\mathcal{O}}^{\hat{T}}$  in equation~(7.19) of \cite{MO95} for $\mathcal{O}=\varepsilon$. Then
\begin{equation}
B_\varepsilon^{T}(d)=-\frac{2\,\pi^{d/2}\,(d-4) (d-2)\, \Gamma
   (d-1)^2\, \Gamma (d)}{\Gamma(3-d/2)\,
   \Gamma(d/2)^4\,
   \Gamma (2 d-3)}
\end{equation}
and $e_0(d)$ follows from
\begin{equation}\label{eq:e0equ}
A_d^{\rm or}\,e_0(d)\,L^{-d}=\frac{B_\varepsilon^{T}(d)}{2^{2-d}}\,\frac{1}{n}\langle T_{zz}(\bm{y},0)\rangle_{L,\rm{c}}=\frac{B_\varepsilon^{T}(d)}{2^{2-d}}\,\Delta_{\rm C} \,\frac{d-1}{L^d},
\end{equation}
where 
\begin{equation}
\langle T_{zz}(\bm{y},0)\rangle_{L,{\rm c}}/n=(d-1)\,\Delta_{\rm C}\,L^{-d}
\end{equation}
is the force per unit area and number $n$ of components at bulk criticality for boundary conditions corresponding to the ordinary transition on both surface planes $z=0$ and $z=L$. The coefficient $\Delta_{\rm C}$ is the Casimir amplitude governing the $L$-dependent contribution $\Delta_{\rm C}L^{1-d}$ of the reduced free energy density per unit area and number $n$ of components at the bulk critical point. (We here and henceforth suppress the subscripts ${}^{\rm or,or}$ at $\Delta_{\rm C}\equiv \Delta_{\rm C}^{\rm or,or}$.)  

It follows from equation~\eqref{eq:e0equ} that
\begin{equation}\label{eq:e0}
e_0(d)=-\frac{2^{d+1}\,(d-1)\pi^{d/2}\,\Gamma[d-1)^2\,\Gamma[d]}{\Gamma(3-d/2)\,\Gamma(d/2)^4\,\Gamma(2d-3)}\,\Delta_{\rm C}.
\end{equation}
The result yields the expansion ($\gamma_E=$ Euler-Mascheroni constant)
\begin{equation}
e_0(4-\epsilon)/\Delta_{\rm C}=-96\pi^2\{1+[2-\gamma_E-\ln(4\pi)]\epsilon/2+\Or(\epsilon^2)\},
\end{equation}
to which the $\epsilon$-expansion result of \cite{EKD93} for the case of the ordinary transition reduces upon setting the factor $(n+2)/(n+8)$ in the latter to its $n=\infty$ limit $1$. For $d=3$, equation~\eqref{eq:e0} yields
\begin{equation}
e_0(3)/\Delta_{\rm C}=-\frac{1024}{\pi}\simeq -325.9493235\ldots.
\end{equation}

\section{Calculation of the coefficient $a_1(3)$}\label{sec:tcorr}

We proceed by computing the coefficient $a_1(d)$. To this end, we will use a strategy analogous to the one followed by Bray and Moore in their calculation of the critical potential \cite{BM77a,BM77c}, i.e.\ of the coefficient $A_d^{\rm or}$. It will be helpful to briefly recall their main steps. Parametrizing  the unknown coefficient as $A_d^{\rm or}=\mu^2-1/4$, they showed that the associated cumulant $\hat{G}(\bm{p};z,z'|\infty,0)$ is given by
\begin{equation}
\hat{G}_\mu(p;z,z')=\sqrt{z_<\,z_>}\,I_\mu(pz_<)\,K_\mu(pz_>),
\end{equation}
where $I_\mu$ and $K_\mu$ are modified Bessel functions of the first and second kind. Further,  $z_<$ and $z_>$ are the smaller and larger of $z$ and $z'$, respectively. 

Substitution of this result into the self-consistency equation~\eqref{eq:sceV} for the critical semi-infinite case $t=0$ and $L=\infty$, with the coupling constant $u$ set to its fixed-point value~\eqref{eq:ustar}, then yielded
\begin{equation}\label{eq:sceVc}
V_d(z|\infty,0)=\frac{u^*\Lambda^{4-d} }{6}\,K_{d-1}\left[J_\mu^{0}(z)-J_\mu^{\Lambda}(z)\right],
\end{equation}
where the functions in square brackets are defined by
\begin{equation}
J_\mu^{\Lambda}(z)=\int_\Lambda^\infty \rmd{p}\,p^{d-2}\left[\hat{G}_\mu(p;z,z)-\frac{1}{2p}\right].
\end{equation}
The first function can be computed in closed form for $2<d<4$ and $z>0$; one obtains
\begin{equation}
J^0_\mu(z>0)=\frac{ \Gamma
   \left(1-d/2\right)
   \Gamma
   \left[(d-1)/2\right]
   \Gamma
   \left[(d-1)/2+\mu
   \right]}{4 \sqrt{\pi }\, \Gamma
   \left[(3-d)/2+\mu\right]\,z^{d-2}}
\end{equation}
Since it gives a $z$-dependence different from $z^{-2}$, the parameter $\mu$ was chosen such that the amplitude $J_\mu^0(1)$ vanishes. This gave the value $\mu=(d-3)/2$ for the case of the ordinary transition. For consistency reasons, the contribution involving $J_\mu^\Lambda(z)$ on the right-hand side of equation~\eqref{eq:sceVc} then must reproduce the critical potential. Using the asymptotic expansions of the Bessel functions in the integrand gives indeed the correct limiting  behaviour
\begin{eqnarray}
J_{\frac{d-3}{2}}^\Lambda(z>0)&=&z^{2-d}\int_{z\Lambda}^\infty\rmd{t}\,t^{d-3}\left[t\,I_{\frac{d-3}{2}}(t)\,K_{\frac{d-3}{2}}(t)-\frac{1}{2}\right]\nonumber\\ 
&\mathop{=}_{z\to\infty}&-\frac{(d-3)^2-1}{16 (4-d) \,z^2}\,\Lambda^{d-4}\left[ 1+O(z^{-2}\Lambda^{-2}) \right]
\end{eqnarray}
for $z\to\infty$, such that the indicated asymptotic term reproduces the critical potential $A_d^{\rm or}z^{-2}$ upon substitution into equation~\eqref{eq:sceVc}. 

Two remarks are in order here. Note, first, that we treated the integrals $J_\mu^0(z)$ and $J_\mu^\Lambda(z)$ as functions of $z$, assuming $z>0$. Owing to its small-$z$ behaviour, the power $z^{-\lambda}$ is not integrable for ${\rm Re}\,\lambda<1$. To obtain well-defined distributions such as $z_+^{-\lambda}$ for such values of $\lambda$ appropriate subtractions at $z=0$ are necessary \cite{GS64}. Specifically, in the (${d=3}$)-dimensional case one can integrate the action of the integrand on the test function $\rme^{-\kappa z}$ over $p$ to see that the distribution $J_0^0(z)$ is proportional to $\delta(z)$. The integrations are straightforward, giving
\begin{eqnarray}
\int_0^\infty \rmd{p}\,p \int_0^\infty\rmd{z}\left[\hat{G}_0(p;z,z)-(2p)^{-1}\right]\rme^{-\kappa z}\nonumber\\ =\int_0^\infty \frac{\rmd p}{p}\left[
\frac{\mathbf{E}\left(\sqrt{1-4 p^2/\kappa^2}\right)-
\mathbf{K}\left(\sqrt{1-4 p^2/\kappa^2}\right)}{4-\kappa^2/p^2}
-\frac{p}{2\kappa}\right]\nonumber \\ =-\frac{1}{4},
\end{eqnarray}
where $\mathbf{K}(k)$ and $\mathbf{E}(k)$ are complete elliptic integrals, which for $k\in (-1,1)$ and $\rmi k\in \mathbb{R}$ can be written as
\begin{eqnarray}
\mathbf{K}(k)=
\int_0^{\pi/2}\frac{d\theta}{\sqrt{1-k^2 \sin^2\theta}},  \quad
\mathbf{E}(k)=\int_0^{\pi/2}{d\theta}\,\sqrt{1-k^2 \sin^2\theta}.
\end{eqnarray}
Thus
\begin{equation}
J_0^0(z)=-\frac{1}{4}\,\delta(z),
\end{equation}
which implies a contribution $-\Lambda\,\delta(z)$ to the non-universal part of $V_3(z|\infty,0)$.

Second, as long as we consider $J^\Lambda_{(d-3)/2}(z)$ and $V_d(z|\infty,0)$ as  functions with $z>0$, we can take the limit $\Lambda\to\infty$. On the other hand, ultraviolet divergences are encountered in integrals for the free energy unless a regularization is used or appropriate subtractions are made.

In order to determine the coefficient $a_1(d)$ we compute the correction linear in $\tau\approx (\ub/6)A_dm^{d-2}$ to the right-hand side of equation~\eqref{eq:sceV} (with $L$ set to $\infty$ and $\ub$ to $u^*\Lambda^{4-d}$) using perturbation theory. Then we match the terms linear in $\tau$ on both sides. This yields
\begin{equation}\label{eq:a1cond}
1-\frac{A_d^{\rm{or}}a_1(d)}{(\Lambda z)^{4-d}}\frac{6}{u^* A_d}=\frac{A_d^{\rm{or}}a_1(d)}{A_d}K_{d-1}\left[H_d(0)-H_d(\Lambda z)\right],
\end{equation}
where $H_d(r)$ is defined as
\begin{equation}\fl
H_d(r)\equiv\int_r^\infty\rmd{s}\int_0^\infty\rmd{\sigma}\,\sigma^{d-3}\left\{I_{\frac{d-3}{2}}[\min(\sigma,s)]\,K_{\frac{d-3}{2}}[\max(\sigma,s)]\right\}^2.
\end{equation}
Upon equating the $z$-independent terms in equation~\eqref{eq:a1cond}, we arrive at
\begin{equation}\label{eq:a1}
a_1(d)=\frac{A_d}{H_d(0)A_d^{\rm{or}}K_{d-1}}.
\end{equation}
For consistency reasons, the term $\propto H_d(\Lambda z)$ must have an asymptotic form that matches with the contribution $\propto (\Lambda z)^{4-d}$ on the left-hand side. 

The required integrations can be done in a straightforward fashion for the $({d=3})$-dimensional case. One obtains
\begin{eqnarray}\label{eq:H3}
\int_0^\infty\rmd{\sigma}\left\{I_0[\min(\sigma,s)]\,K_0[\max(\sigma,s)]\right\}^2\nonumber \\ =
s K_0(s){}^2 \,
   _2F_3\left(\frac{1}{2},\frac{1}{2};1,1,\frac{3}{2};s^2\right)
+\frac{1}{4} \sqrt{\pi }
   I_0(s){}^2\,
   G_{2,4}^{4,0}\left(s^2\bigg|
\begin{array}{c}
 1,1 \\
   0,\frac{1}{2},\frac{1}{2},\frac{1}{2} \\
\end{array}
\right)\nonumber\\
=\frac{1}{4s^2}+O(s^{-4}),
\end{eqnarray}
where ${}_pF_q(a_1,\ldots,a_p;b_1,\ldots,b_q;x)$  and $G_{p,q}^{m,n}\Big(x\Big|
\begin{array}{c}
 a_1,\ldots,a_p \\
   b_1,\ldots,b_q
\end{array}\Big)$ denote the generalised hypergeometric and the Meijer $G$-function, respectively.
Performing the integration $\int_0^\infty\rmd{s}$ of the result displayed in the second line gives
\begin{equation}
H_3(0)=\frac{\pi^2}{8},
\end{equation}
from which the value
\begin{equation}
a_1(3)=-\frac{16}{\pi^2}
\end{equation}
follows via equation~\eqref{eq:a1}. Integration of the expansion shown in the last line of equation~\eqref{eq:H3}
yields 
\begin{equation}
\frac{u^*(3)}{6}K_2\,H_3(\Lambda z)\mathop{=}_{\Lambda z\to\infty}\frac{1}{\Lambda z}+\ldots.
\end{equation}
Hence the consistency condition is satisfied.

Using {\sc Mathematica}\footnote{Wolfram Research, Computer code {\sc Mathematica}, version 9.}
the integral $\int_0^\infty\rmd{s}$ required for $H_d(r)$ with $2<d<4$ can be determined in terms of generalised hypergeometric functions. However, we did  not manage to obtain a closed analytic expression for $H_d(r)$ or $H_d(0)$ for $3\ne d\in (2,4)$.  Since our main interest is in the ($d=3$)-dimensional case, we refrain from making further efforts to compute these integrals in closed analytical form.

\section{Consequences}\label{sec:conseq}
\subsection{Scaling forms of finite-size and excess free energies and  Casimir force}

We now turn to the question what consequences the above exact results have for surface and finite-size critical behaviour. To this end we introduce the partition function
$
 \mathcal{Z}=\int\mathcal{D}[\bm{\phi}]\,\rme^{-\mathcal{H}[\bm{\phi}]}
$
and the $n\to\infty$ limit of the reduced free energy of the slab per area  $A=\int_{\mathbb{R}^{d-1}}\rmd^{d-1}y$ and number $n$ of components,
\begin{equation}
f_L=-\lim_{n\to\infty}\frac{\ln\mathcal{Z}}{n A},
\end{equation}
where we assume that $\mathcal{Z}$ is computed subject to boundary conditions appropriate for the ordinary transition. This quantity can be decomposed  as \begin{equation}
f_L(T,L)=Lf_{\rm{b}}(T)+2\,f_{\rm{s}}(T)+f_{\rm{res}}(T,L)
\end{equation}
into a contribution proportional to the bulk free energy density $f_{\rm{b}}(T)$, an $L$-independent surface contribution $f_{\rm{s}}(T)$ (the reduced surface excess free energy), and the $L$-dependent reduced residual free energy $f_{\rm{res}}(T,L)$. On large length scales, the latter  takes the scaling form
\begin{equation}\label{eq:fresscalf}
f_{\rm{res}}(T,L)\approx L^{-(d-1)}\,\Theta_d[t(\Lambda L)^{1/\nu}].
\end{equation}

Differentiating $f_{\rm{res}}(T,L)$ with respect to $L$, we can define a ``Casimir'' force $\mathcal{F}_{\rm{C}}$ via
\begin{equation}
\beta\mathcal{F}_{\rm{C}}(T,L)\equiv -\frac{\partial}{\partial L}f_{\rm{res}}(T,L)
\end{equation}
where $\beta=(k_{\rm{B}}T)^{-1}$. Equation~\eqref{eq:fresscalf} yields the scaling form
\begin{equation}
\beta\mathcal{F}_{\rm{C}}(T,L)\approx L^{-d}\,\vartheta_d[t(\Lambda L)^{1/\nu}].
\end{equation}
with
\begin{equation}
\vartheta_d(x)=(d-1)\,\Theta_d(x)-\frac{x}{\nu}\,\Theta_d'(x).
\end{equation}

The bulk and surface free energies $f_{\rm{b}}$ and $f_{\rm{s}}$  can be decomposed into regular and singular parts:
\begin{eqnarray}
f_{\rm{b}}(T)&=&f^{\rm{sing}}_{\rm{b}}(t)+f^{\rm{reg}}_{\rm{b}}(T)=f^{\rm{sing}}_{\rm{b}}(t)+\Lambda^d\sum_{k=0}^\infty f_k\,t^k,\nonumber\\
f_{\rm{s}}(T)&=&f^{\rm{sing}}_{\rm{s}}(t)+f^{\rm{reg}}_{\rm{s}}(T)=f^{\rm{sing}}_{\rm{s}}(t)+\Lambda^{d-1}\sum_{k=0}^\infty f^{(\rm{s})}_k\,t^k.
\end{eqnarray}
The leading singular part of $f_{\rm{b}}$ may be gleaned  from the literature and is computed within the framework of our cutoff regularization scheme in the appendix. The result is
\begin{equation}\label{eq:fbsing}
f^{\rm{sing}}_{\rm{b}}(t)/\Lambda^d\approx \frac{d-2}{2d}\,A_d\,|t|^{d/(d-2)}\,\theta(t)\mathop{=}_{d=3}\frac{1}{24\pi}\,t^3\,\theta(t),
\end{equation}
where $\theta(t)$ is the Heaviside theta function.
The singular part of $f_{\rm{s}}$ is known to vary $\sim |t|^{2-\alpha_{\rm{s}}}$ with $\alpha_{\rm{s}}=\alpha+\nu=(d-1)/(d-2)$ \cite{Die86a,Die97,Bin83}. Hence we write
\begin{equation}
f^{\rm{sing}}_{\rm{s}}(t)/\Lambda^{d-1}\approx A^{(\rm{s})}_\pm(d) \,|t|^{(d-1)/(d-2)}.
\end{equation}

\subsection{Regularity requirements and implied singularities of the scaling function $\Theta$}
Since the exponent $2-\alpha_{\rm{s}}$ takes the integer value $2$ at $d=3$, the latter singular term interferes with the regular one $\propto t^2$. Thus the amplitudes $A^{(\rm{s})}_\pm(d)$ may be expected to have simple poles at $d=3$. Let us first consider the case $2<d<4$ with $d\ne 3$. Whenever the thickness $L<\infty$, no phase transition occurs at $t=0$. Consequently, the free energy density $f_L$ must be regular at $t=0$. The consequences of this regularity constraint on scaling functions were previously discussed in the literature (see e.g.\ \cite[Sec.~VII]{KD92a} and \cite[Sec.~4]{GD08}). The scaling function $\Theta(tL)$ must have singularities that cancel those of the contributions $L f^{\rm{sing}}_{\rm{b}}(t)$ and $f^{\rm{sing}}_{\rm{s}}(t)$. Thus $\Theta(x)$ must behave as
\begin{equation}
\Theta_{d\ne3}(x)\mathop{=}_{x\to 0\pm}-A_\pm(d)\,|x|^{\frac{d}{d-2}}-2A_\pm^{(\rm{s})}(d)\,|x|^{\frac{d-1}{d-2}}+\Delta_{\rm{C}}+\ldots,
\end{equation}
where the ellipsis stands for terms regular in $x$ of first and higher orders in $x$. 

The case of $d=3$ requires special consideration. By analogy with equation~\eqref{eq:b0exp}, we anticipate the Laurent expansions
\begin{equation}
A^{(\rm{s})}_\pm(d)=-\frac{f^{(\rm{s})}_{2,-1}}{d-3}+A^{(\rm{s})}_{0,\pm}-f^{(\rm{s})}_{2,0}+\Or(d-3)
\end{equation}
and
\begin{equation}
f_2^{(\rm{s})}(d)=\frac{f^{(\rm{s})}_{2,-1}}{d-3}+f^{(\rm{s})}_{2,0}+\Or(d-3)
\end{equation}
for the amplitudes $A^{(\rm{s})}_\pm(d)$ and the coefficient $f_2^{(\rm{s})}(d)$. This implies the limit
\begin{equation}
\lim_{d\to 3}\left[A^{(\rm{s})}_\pm(d)|t|^{\frac{d-1}{d-2}}+f^{(\rm{s})}_2(d)\,t^2\right]=t^2\left[A_{0,\pm}^{(\rm{s})}-f^{(\rm{s})}_{2,-1}\ln |t|\right].
\end{equation}
On the other hand, we know that $A_\pm(d)$ does not have a pole at $d=3$. Using this in conjunction with the result given in equation~\eqref{eq:fbsing}, we can conclude that the $t^3$ contributions to $f_{\rm{b}}$ add up to
\begin{equation}
\left[A_\pm(d)\,|t|^3+f_3\,t^3\right]_{d=3}=\Lambda^3 \left[\frac{1}{24\pi}\theta(t)+f_3\right]t^3
\end{equation}
when $d=3$. To comply with the value of \cite{AH75} of the universal amplitude ratio given in equation~\eqref{eq:ApAm}, we would have to choose $f_3=-(6\pi^3)^{-1}$. Modifying the leading singular part of $f_{\rm{b}}^{\rm{sing}}$ in this fashion would imply a corresponding change of the scaling function $\Theta(x)$ so that the contributions $\propto f_3$ cancel out in $f_L$. Hence we can just as well set $f_3=0$ and continue to work with the result for $A_\pm(3)$ stated in equation~\eqref{eq:fbsing}. It follows that the scaling function $\Theta_3$ behaves as
\begin{equation}\fl
\Theta_3(x)=\Delta_C-\left[A_{0,+}^{(\rm{s})}-A_{0,-}^{(\rm{s})}+\frac{x}{48\pi}\right]2x^2\, \theta(x)%
+2f^{(\rm{s})}_{2,-1}\,x^2\ln|x|+\ldots,
\end{equation}
where the ellipsis denotes again contributions regular in $x$ of at least linear order. Since the scaling function $\Theta(x)$ is universal, so must be the difference
\begin{equation}
\Delta A^{(\rm{s})}_0=A_{0,+}^{(\rm{s})}-A_{0,-}^{(\rm{s})}
\end{equation}
and the amplitude $f^{(\rm{s})}_{2,-1}$. By contrast, the coefficients $A_{0,\pm}^{(\rm{s})}$ are not universal, just as the amplitudes $A_{0,\pm}^{(\rm{s})}$ are not.\footnote{The universality of $\Delta A_0^{(\rm{s})}$ follows also from the universality of $f^{(\rm{s})}_{2,-1}$ and the ratio $A^{(\rm{s})}_+/A^{(\rm{s})}_-=\Delta A^{(\rm{s})}_0/f^{(\rm{s})}_{2,-1}+\Or(d-3)$.}

We now show that the coefficient $f_{2,-1}^{(\rm{s})}$ is given by
\begin{equation}\label{eq:f2m1}
f_{2,-1}^{(\rm{s})}=-\frac{a_1(3)}{64\pi}=\frac{1}{4\pi^3}.
\end{equation}

To this end, we use the known $n=\infty$ expression \cite{MZ03,DGHHRS12}
\begin{eqnarray}\label{eq:flexp}
f_L&=&\frac{1}{2}\int_0^L\rmd{z}\bigg\{K_{d-1}\int_0^\Lambda\rmd{p}\,p^{d-2}\,\langle z|\ln(p^2-\partial_z^2+V_d)|z\rangle\nonumber\\&&\strut -\frac{3}{\ub}(\tb_c+\tau-V_d)^2\bigg\}+f_L^{(0)},
\end{eqnarray}
where $f_L^{(0)}$ is a regular background term which does not matter henceforth. If we subtract from the integrand its value at $z=L=\infty$, we can set $L=\infty$ to obtain an equation for the excess surface free energy $f_{\rm{s}}$,. This we differentiate with respect to $\tau$, taking into account the self-consistency condition~\eqref{eq:sceV} and the relation~\eqref{eq:Vbulk} for the bulk value of $V_d$. The excess energy density thus becomes
\begin{equation}
e_{\rm{s}}(\tau)\equiv\frac{\partial}{\partial \tau}f_{\rm{s}}=\frac{3 }{u}\Lambda^{d-4}\int_0^\infty\rmd{z}\left[V_d(z|\infty,\tau)-m^2\right]
\end{equation}
We subtract the value of this quantity at criticality and decompose the integral as $\int_{z_{\rm{min}}}^{1/m}\rmd{z}+\int_{1/m}^\infty\rmd{z}$ into contributions from the boundary and inner regions $z_{\rm{min}}\le z\le 1/m$ and $1/m\le z<\infty$, where $z_{\rm{min}}$ is a short-distance cutoff $\sim \Lambda^{-1}$ needed  for convergence. In the boundary region we may use the results of the BOE expansion given in equations~\eqref{eq:BOEres}--\eqref{eq:BOEYd2} to approximate the integrand. In the inner region, the integrand vanishes $\propto \exp(-2mz)$ as $z\to\infty$. The dominant contribution in the boundary region is the one proportional to $a_1(d)$. Depending on whether $d=3$ or $3<d<4$, it gives rise to a $t\ln t$ singularity or causes the amplitude of the  thermal singularity $\sim t^{1/(d-2)}$ to have a pole at $d=3$. One finds
\begin{eqnarray}\fl
\Lambda^{3-d}\left[e_{\rm{s}}(\tau)-e_{\rm{s}}(0)\right]&=&\frac{3}{u\Lambda}\,\int_{z_{\rm{min}}}^{1/m}\rmd{z}A^{\rm{or}}_dz^{-2}\left[a_1(d)\,(mz)^{d-2}+\ldots\right]+\ldots\nonumber\\
&\approx&\frac{3t}{u}\cases{\frac{t^{(3-d)/(d-2)}}{d-3}\left[\frac{-1}{4}a_1(3)+\Or(d-3)\right]& for $3<d<4$,\\
\frac{1}{4}a_1(3)\,\ln(t\Lambda z_{\rm{min}})&for $d=3$.
}
\end{eqnarray}
The residue $f_{2,-1}$ can be found from either result by matching the given singularities with those of $(\partial t/\partial\tau)\,\partial_t f_{\rm{s}}^{\rm{sing}}$ using the relation~\eqref{eq:tdef}. One thus obtains equation~\eqref{eq:f2m1}. 

\section{Summary of exact results and concluding remarks}\label{sec:conc}

It will be helpful to summarise the exact results obtained above and their consequences briefly, focusing on the $({d=3})$-dimensional case.
We have extended Bray and Moore's exact result \cite{BM77a,BM77c} for the semi-infinite critical case to temperatures away from $T_{\rm c}$ by determining the leading temperature correction to the self-consistent potential $V_3(z|{L=\infty},\tau)$ . The result can be written as
\begin{equation}\label{eq:Vnsb}
V_3[z|\infty,\tau(t)]\mathop{=}_{z\to 0}-\frac{1}{4z^2}+\frac{4\,\rm{sgn}(t)}{\pi^2z\,\xi(|t|)}+\Or(z^0),
\end{equation}
where $\rm{sgn}(t)$ means the sign of the temperature variable $t\propto (T-T_{\rm c})/T_{\rm c}$ and $\xi(|t|)$ is the bulk correlation length in the disordered phase. An analogous near-surface behaviour holds near the surface plane at $z=L$.

Using the result~\eqref{eq:Vnsb} enabled us to determine the leading temperature singularity of the excess surface free energy density $f_{\rm s}$. It can be written as
\begin{equation}
f_{\rm{s}}^{\rm{sing}}\big|_{d=3}\mathop{\approx}_{t\to 0} -\frac{1}{4\pi^3}\,[\xi(|t|)]^{-2}\ln[\xi(1)/\xi(|t|)].
\end{equation}
Here the argument of the logarithm is precisely the absolute value $|t|$ of the temperature variable $t$. 

We then exploited the condition that the total free energy density $f_L$ must be an analytic function of $t$ near  $t=0$ to determine the behaviour of the scaling function 
\begin{equation}\label{eq:Thetasing}
\Theta_3(x)\mathop{\approx}_{x\to 0} \Delta_{\rm{C}}-\left[\Delta A_0^{(\rm{s})}+\frac{x}{48\pi}\right]2x^2\,\theta(x)+\frac{1}{2\pi^3}\,x^2\ln|x|+\ldots
\end{equation}
of the residual free energy $f_{\rm res}(T,L)$ as a function of the scaling argument $x=\rm{sgn}(t)\,L/\xi(|t|)$. Here the ellipsis stands for regular and less singular terms. Our knowledge of the $t$-dependence of $V(z|\infty,\tau)$ obtained here is not sufficient to determine the universal difference $\Delta A_0^{(\rm{s})}$. In a forthcoming paper \cite{RD14} we will show how this quantity can be obtained via inverse scattering methods.

The asymptotic form  of $\Theta_3$ given in equation~\eqref{eq:Thetasing} implies that the scaling function $\vartheta_3$ of the critical Casimir force behaves as
\begin{equation}
\vartheta_3(x)\mathop{\approx}_{x\to 0}\frac{x^3}{24\pi}\theta(x)-\frac{x^2}{2\pi^3}+2\Delta_{\rm{C}}+\ldots.
\end{equation}
An immediate consequence is that the second derivative of this function at the critical point takes the value
\begin{equation}
\vartheta''_3(0)=-\pi^{-3}.
\end{equation}
High-precision results obtained by means of numerical solutions of the  ${n=\infty}$ equations \cite{DGHHRS13} confirm the latter analytical value.

Finally, we exploited results gleaned from the literature \cite{MO95} to determine the distant-wall correction of the critical potential, obtaining
\begin{equation}
V_3(z|L,0)\mathop{\approx}_{z/L\to 0}-\frac{1}{4z^2}\left(1-\frac{1024}{\pi}\,\Delta_{\rm{C}}\,z^3L^{-3}\right).
\end{equation} 
The numerical results of the paper  \cite{DGHHRS13} mentioned above accurately agree also with this prediction.

Exact analytical results such as those summarised above provide useful benchmarks for numerical investigations and analytical approximations. It would certainly be desirable to determine the exact solution to the self-consistency $n=\infty$ equations in closed analytical form. As can be seen from our above analysis, this is a highly nontrivial task. Some progress in this direction can be made with the help of inverse scattering and other methods \cite{RD14,DGHHRS13}, but the challenge to obtain an exact analytical solution remains.

\ack We thank Martin Hasenbusch, Alfred Hucht and Felix Schmidt for fruitful interactions and discussions. Partial support of this work by Deutsche Forschungsgemeinschaft (DFG), in its initial phase via grant Di 378/5 and subsequently by grant Ru 1506/1, is also gratefully acknowledged.
\appendix
\section{Calculation of the bulk free energy density}

The singular part of the bulk free energy density $f_{\rm{b}}$ can be inferred from results available in the literature. However, since non-universal metric factors depend also on the chosen cutoff regularization, one must choose the scales of the scaling fields properly to translate results based on a different cutoff scheme to ours. For convenience, we here give a brief derivation based on our scheme of restricting integrations over parallel momenta $\bm{p}$  to magnitudes $p\le \Lambda$. It will be  sufficient to do the calculation for the disordered phase $t>0$. In the ordered bulk phase $t<0$, the inverse transverse susceptibility $m_T^2 $ vanishes on the coexistence curve. According to \cite{MZ03}, the leading singular part of $f_{\rm{b}}$ for $t<0$ has the same form as for $t>0$, except that $m$ must be replaced by $m_T$. Hence it vanishes.

It is convenient to add a constant to the Hamiltonian such that $f_{\rm{b}}|_{T_{\rm{c}}}=0$. Then we have for $\tau>0$ (see e.g.\ \cite{MZ03,DGHHRS12})
\begin{equation}\fl
f_{\rm{b}}=\frac{1}{2}\int_0^{m^2}\rmd{r}\int_0^\infty\frac{\rmd{k}}{\pi}K_{d-1}\int_0^\Lambda \rmd{p}\,p^{d-2}\frac{1}{p^2+k^2+r}-\frac{3}{2\ub}[(\tb_{\rm{c}}+\tau-m^2)^2-\tb_{\rm{c}}^2]
\end{equation}
where $K_{d-1}$ is defined in equation~\eqref{eq:Kd}.
The integrations can be performed in a straightforward fashion to obtain
\begin{eqnarray}\fl
f_{\rm{b}}(\tau\ge 0)=\frac{ {}_2F_1[-1/2,(d-1)/2;(d+1)/2;-\Lambda^2/m^2]\,m/\Lambda-1 +1/d}{2^d \pi^{(d-1)/2}\, \Gamma[(d+1)/2]}\,\Lambda^d\nonumber\\
\fl\phantom{f_{\rm{b}}(\tau\ge 0)=} \strut -\frac{3}{2\ub}(m^4-2m^2\tau+\tau^2-2m^2\tb_{\rm{c}}+2\tau\tb_{\rm{c}}).
\end{eqnarray}

To determine the leading thermal singularity, we expand in $m$ using the asymptotic $m$-dependence of $\tau$ given in equation~\eqref{eq:taum}. The leading singular terms are those proportional to $m^d$ and to $m^2\tau$. We thus arrive at
\begin{equation}
f^{\rm{sing}}_{\rm{b}}(\tau)\approx \theta(\tau)\left[- \frac{A_d}{d}\,m^d+\frac{3m^2\,\tau}{\ub}\right].
\end{equation}
Expressing the right-hand side in terms of the temperature variable $t$ introduced in equation~\eqref{eq:tdef} then leads to the result given in equation~\eqref{eq:fbsing}.

\section*{References}
%

\end{document}